\documentclass[aps,pra,reprint]{revtex4-1}

\usepackage{gensymb}
\usepackage{amsmath}
\usepackage{amssymb}
\usepackage{bm}
\usepackage{color}
\usepackage[linkcolor=blue, colorlinks=true, urlcolor=blue,
citecolor=red]{hyperref}
\usepackage{graphicx}
\usepackage[caption=false]{subfig}
\usepackage{bbold}
\usepackage{verbatim}
\usepackage{appendix}
\usepackage{amsthm}
\usepackage[normalem]{ulem}

\theoremstyle{plain}

\newcommand{\ket}[1]{\left| #1 \right>} 
\newcommand{\bigket}[1]{\big| #1 \big\rangle} 
\newcommand{\bra}[1]{\left< #1 \right|} 
\newcommand{\braket}[2]{\left< #1 \vphantom{#2} \right|
	\left. #2 \vphantom{#1} \right>} 
\newcommand{\bigbraket}[2]{\big\langle #1 \vphantom{#2} \big|
	 #2 \vphantom{#1} \big\rangle} 
\newcommand{\ketbra}[2]{\ket{#1}\hspace*{-0.mm}\bra{#2}}

\newcommand{\Eq}[1]{Eq.~\eqref{#1}}
\newcommand{\av}[1]{\left\langle #1 \right\rangle}
\newcommand{\bigav}[1]{\big\langle #1 \big\rangle}
\newcommand{\Id}{\mathbb{1}}
\newcommand{\Tr}[1]{\mathrm{\text{Tr}}\left[#1\right]}
\newcommand{\nop}{\hat{n}}
\newcommand{\Iop}{\hat{I}}
\newcommand{\aop}{\hat{a}}
\newcommand{\bop}{\hat{b}}

\newcommand{\bE}{{\bf E}}
\newcommand{\bepsilon}{{\pmb \epsilon}}

\newcommand{\bt}{{\bf t}}

\begin{document}

\title {Optical nonclassicality test based on third-order intensity correlations}
\date{\today}

\author{L. Rigovacca$^1$}
\author{W. S. Kolthammer$^1$}
\author{C. Di Franco$^{2,3}$}
\author{M. S. Kim$^1$}
\affiliation{$^1$QOLS, Blackett Laboratory, Imperial College London, London, SW7 2AZ, United Kingdom}
\affiliation{$^2$School of Physical and Mathematical Sciences, Nanyang Technological University, 637371, Singapore}
\affiliation{$^3$Complexity Institute, Nanyang Technological University, 637723, Singapore}

\begin{abstract}
		We develop a nonclassicality criterion for the interference of three delayed, but otherwise identical, light fields in a three-mode Bell interferometer. We do so by comparing the prediction of quantum mechanics with those of a classical framework in which independent sources emit electric fields with random phases. In particular, we evaluate third-order correlations among output intensities as a function of the delays, and show how the presence of a correlation revival for small delays cannot be explained by the classical model of light. The observation of a revival is thus a nonclassicality signature, which can be achieved only by sources with a photon-number statistics that is highly sub-Poissonian.
		Our analysis provides strong evidence for the nonclassicality of the experiment discussed by Menssen \emph{et al.}, [A. J. Menssen \emph{et al.}, Phys. Rev. Lett. {\bf 118}, 153603 (2017)], and shows how a collective ``triad'' phase affects the interference of any three or more light fields, irrespective of their quantum or classical character. 
\end{abstract}

\maketitle

\section{Introduction}
Thanks to recent experimental advancements in photonics, the possibility of creating and manipulating multiple single-photon states has now become a reality \cite{Spring_17,Tillmann2015,Rome_experiment,Oxford_experiment,Crespi2013,Spring2013,Yao2012,Matthews2011,Wang_2017},
and even the genuine interference of three single photons has been recently isolated and observed \cite{Alex_2017,Jennewein_2017}. 
More generally, the development of on-chip wave-guides, sources, and detectors, can be used to test quantum mechanical predictions in previously unavailable regimes. For example, this can be done by pursuing fundamental research on boson sampling \cite{Aaronson2011,Shchesnovich2016,Bentivegna2015,Broome2013,Crespi2013,Spring2013,Spagnolo2014} or on the effects of many-particle interference \cite{Tichy2012,Belinskii1992,Lim2005,Mayer2011,Laibacher2015,Tichy2011,Ra2013,Tichy2012,Tichy2011,Ra2013,Spagnolo_multiInt_2014,Tamma2015}. Rather than considering the probability distribution of output detection events, it is often convenient to study the correlations among detected output intensities. In particular, correlations among pairs of intensities have been used to acquire information on the nature of particles scattering through a boson sampling setup \cite{Walschaers2016BS,Walschaers2016}, and also to develop nonclassicality criteria in multiport linear optical interferometers \cite{Rigovacca_2016}. 
On the other hand, the correlations among three or more output intensities have been studied in the context of imaging resolution \cite{HighOrder_1,HighOrder_2,HighOrder_3,HighOrder_4}, or to study signatures of many-particle interferences in quantum walks \cite{Mayer2011}.

Although the possibility of detecting the nonclassicality of light by studying correlation functions of photon number operators has been previously explored \cite{Shchukin2005, Nori_2010, Sperling_2012, Sperling_SubBinomial_2012, Sperling2013, Sperling2015, Sperling2017, SperlingPRL_17, Lachman_2016, Moreva_2017}, in this paper we take a different approach tailored to the evolution of multimode light fields through linear optical interferometers, similarly to what has been done in \cite{Rigovacca_2016}. In particular, we first choose a linear optical interferometer and the input light fields in such a way that certain hypotheses, detailed and commented in the following, are satisfied. Then, by having access only to the output fields of the interferometer, our aim is to identify measurement outcomes that cannot be explained by means of a classical wave-like model of light. As a consequence, as long as the aforementioned hypotheses are satisfied, the observation of such outcomes would imply the nonclassicality of the input light.

In this paper, we consider the interference of three independent quantum light fields diagonal in the Fock basis through a three-mode linear optical setup. In particular, we discuss the quantum features that can be observed in the third-order correlation function
\begin{equation}\label{def: G3 I quantum}
G_3^{(\text{Q})} = \frac{\bigav{\Iop_1 \Iop_2 \Iop_3}}{\bigav{\Iop_1} \bigav{\Iop_2} \bigav{\Iop_3}},
\end{equation} 
where $\{\Iop_i\}_{i=1}^3$ are the operators associated with the output intensities, by comparison with its classical counterpart
\begin{equation}\label{def: classical I G3}
G_3^{(\text{cl})} = \frac{\av{I_1 I_2 I_3}}{\av{I_1} \av{I_2} \av{I_3}}.
\end{equation}
In this latter case, we substitute quantum states with corresponding classical light fields emitted with random phases by independent stochastic sources. The use of independent classical fields characterized by random phases, and the corresponding quantum product states diagonal in the Fock basis, represent the core hypothesis required for our nonclassicality test. As these conditions are easily tested, and naturally arise in optical experiments involving independent sources, the presence of these assumptions does not significantly limit the generality of our result. Moreover, the same framework was used in the past \cite{Mandel_ClBound,Paul_ClBound,HOM_book,Ou_interference} in order to prove the nonclassicality of the well-known Hong-Ou-Mandel effect~\cite{HOM_PRL_1987}.

A common approach to testing nonclassicality, as in Ref.~\cite{Rigovacca_2016}, is to find values of $G_3^{(\text{Q})}$ that are unobtainable by $G_3^{(\text{cl})}$. Instead, motivated by generalized interference ``dips'' recently observed in three-photon experiments \cite{Spring_17,Tillmann2015,Rome_experiment,Alex_2017}, we consider the ``shape'' of the interference patterns. In particular, we study interference when two out of three identical input fields, which are assumed to have Gaussian wave packets in time with variance $\sigma^2$, are respectively delayed by $\pm\tau$. 
More precisely, our nonclassicality criterion is based on the observation that for a Bell three-port interferometer the function $G_3^{(\text{cl})}(\delta)$ always takes the shape of a regular dip, with respect to the dimensionless delay $\delta= \sigma\tau$. Instead, under the same conditions quantum sources can lead to a ``revival'' in $G_3^{(\text{Q})}(\delta)$ when $\delta$ approaches zero, which can thus be considered a quantum signature. Its origin will be shown to depend on the presence of input quantum states with a photon-number statistics that is highly sub-Poissonian. 
This means that, unlike Ref.~\cite{Rigovacca_2016}, not any sub-Poissonian quantum state can exhibit this indicator of nonclassicality. Despite this limitation on the set of testable nonclassical states, the approach presented in this paper has the advantage of being easier to implement experimentally, at least for single-photon sources. In this case, $G_3^{(\text{Q})}$ corresponds to the probability of coincident events, in which a photon is detected from each output spatial mode.

For general linear optical interferometers, it is possible to observe a revival in $G_3^{(\text{cl})}$ for classical sources that emit light with non-uniform intensities. However, this is not expected for interferometers that are ``close'' enough to the Bell interferometer, a conjecture we support with numerical analysis. It is therefore possible to apply our nonclassicality test in realistic conditions, where the interferometric setup cannot be fixed with arbitrary precision.
For example, our result strongly challenges the possibility of explaining the experimental data of Ref.~\cite{Alex_2017} by means of the aforementioned classical wave-like model of light. This is because in that experiment three single-photons have been injected in a Bell three-port interferometer, and a revival in the probability of coincident events has been observed. 
Our analysis also provides insight into the ``triad phase'' introduced in Ref.~\cite{Alex_2017} to describe an aspect of particle distinguishability that arises only when three or more independent particles are involved. We show that a corresponding phase arises in the interference of three or more fields. In this context, the triad phase is shown to be independent of the quantum character of the fields.

The remainder of this paper is organized as follows. In Sec.~\ref{sec: g3 functions} we formally introduce the classical wave-like model of light that will be used throughout the paper, as well as its quantum counterpart. In both cases we obtain explicit expressions for the third-order correlation function $G_3$, and a first comparison between the two is performed. 
The idea of ``revival'' will then be introduced in Sec.~\ref{sec: revivals}, by considering a specific example in which three single-photon sources, or three classical fields with the same intensities, evolve through a Bell interferometer. 
The core of our nonclassicality test will be developed in Sec.~\ref{sec: norev results}, where we show that a revival cannot appear classically for linear optical setups that are equal to, or close to, a Bell three-port interferometer. In Sec.~\ref{sec: nonclassicality}, instead, we repeat a similar analysis in a quantum setting, with the purpose of identifying which features of the input quantum states are responsible for the appearance of a revival.
A final discussion and our conclusions are left for Sec.~\ref{sec: conclusion}.

\section{Third-order intensity correlation functions} \label{sec: g3 functions}
In this section we provide formal expressions for the third-order correlation functions $G_3^{(\text{cl})}$ and $G_3^{(\text{Q})}$, when the light emitted by independent sources evolves through a three-mode linear optical interferometer. We will first focus on classical fields characterized by random phases, and then on quantum states with density matrices that are diagonal in the Fock basis.
In both scenarios we assume that each source emits light in a fixed frequency-polarization mode of light, characterizing the distribution of each emitted wave-packet in frequency and polarization.
This analysis allows us to discuss the similarities and the differences between the classical and the quantum expressions. In particular, we show how the triad phase identified and studied in Ref.~\cite{Alex_2017} naturally emerges also in a classical framework whenever three or more sources interfere.

\subsection{Classical scenario}\label{sec: classical scenario}
The classical model of light discussed in this paper consists of three stochastic independent sources emitting pulses of light characterized by random phases and possibly varying intensities. Formally, the electric field $\bE_\alpha(t)$ with phase $\xi_\alpha$ and intensity $I_\alpha$ is emitted by the $\alpha$th source with probability $p_\alpha(\xi_\alpha,I_\alpha) = \tfrac{1}{2\pi}p_\alpha(I_\alpha)$, and can be expanded as
\begin{equation}\label{def: classical field}
\bE_\alpha (t) = e^{i\xi_\alpha}\sqrt{I_\alpha} \sum_{\zeta = 1}^2 \int_{0}^\infty \text{d}\omega \,\phi_\alpha(\omega,\zeta) \frac{e^{-i \omega t}}{\sqrt{2\pi}} \bepsilon_{\omega,\zeta}.
\end{equation}
Here $\bepsilon_{\omega,\zeta}$ is the unit vector that characterizes the polarization associated with the angular frequency $\omega$, satisfying $\bepsilon_{\omega,\zeta}\cdot\bepsilon_{\omega,\zeta^\prime}^* = \delta_{\zeta\zeta^\prime}$. The total intensity of the $\alpha$th source is  
\begin{equation}\label{eq: source intensity}
I_\alpha = \sum_{\zeta = 1}^2 \int_{-\infty}^{+\infty} \text{d}t \, \bE^*_\alpha (t) \cdot \bE_\alpha (t),
\end{equation}
and $\phi(\omega,\zeta)$ is a normalized frequency-polarization mode function in $L^2(\mathbb{R})\otimes \mathbb{C}^2$, i.e., such that 
\begin{equation}
\sum_{\zeta = 1}^2 \int_0^\infty \text{d}\omega \, |\phi(\omega,\zeta)|^2 = 1.
\end{equation}
Notice that $\phi_\alpha(\omega,\zeta)$ is defined up to a global phase that could be included in $\xi_\alpha$. As a consequence, it can be represented by a vector $\ket{\phi_\alpha}$ in Dirac notation with components
\begin{equation}
\ket{\phi_\alpha} = \sum_{\zeta = 1}^2 \int_{0}^{+\infty} \text{d}\omega \, \phi_\alpha(\omega,\zeta) \; \ket{\omega,\zeta},
\end{equation}
where $\{\ket{\omega,\zeta}\}_{\omega,\zeta}$ is an orthonormal basis for $L^2(\mathbb{R})\otimes \mathbb{C}^2$.

After the light has been emitted by the sources, it evolves through a linear optical interferometer. Labeling the ports of the interferometer as inputs ($\alpha$, $\beta$, $\gamma$) and outputs ($i$, $j$, $k$), we can write the evolution of the electric fields as
\begin{equation}\label{eq: classical interferometer}
\bE_i (t) = \sum_{\alpha=1}^{3} U_{i\alpha}  \bE_\alpha (t),
\end{equation}
for $i = 1,2,3$, where $U$ is the matrix characterizing the linear optical setup. With this notation, the quantities $\{I_i\}_{i=1}^3$ that appear in $G_3^{(\text{cl})}$ [see \Eq{def: classical I G3}] are the integrated output intensities of the fields, formally obtained as in \Eq{eq: source intensity} by integrating the moduli of the output electric fields. 
The expectation value $\av{\cdot}$ appearing in \Eq{def: classical I G3}, instead, represents the average over all possible realizations of the sources, i.e., an average weighted by the probability
\begin{equation}
p^{(\text{cl})} \left(\{\xi_\alpha\}_\alpha,\{I_\alpha\}_\alpha\right) = \prod_{\alpha=1}^{3} \frac{1}{2\pi} p_\alpha(I_\alpha).
\end{equation}

When evaluating correlation functions among output intensities, one computes several integrals of the form
\begin{align}\label{eq: classical fields overlap}
I_j &= \sum_{\alpha,\beta=1}^{3} U_{j\alpha}^*U_{j\beta} 
\sum_{\zeta = 1}^2 \int_{-\infty}^{+\infty} \text{d}t \, \bE^*_\alpha (t) \cdot \bE_\beta (t) \notag\\
& = \sum_{\alpha\beta=1}^{3} U_{j\alpha}^*U_{j\beta} \sqrt{I_\alpha I_\beta} \; \bigbraket{\phi_\alpha}{\phi_\beta} e^{i(\xi_\beta - \xi_\alpha)},
\end{align} 
where $\bigbraket{\phi_\alpha}{\phi_\beta}$ characterizes the distinguishability of the fields coming from sources $\alpha$ and $\beta$. In particular, it is convenient to write these overlaps as
\begin{equation}\label{eq: classical overlaps}
\bigbraket{\phi_\alpha}{\phi_\beta} = r_{\alpha\beta} \, e^{i \psi_{\alpha\beta}},
\end{equation}
with $0 \leq r_{\alpha\beta} \leq 1$. 
Note that for $\alpha=\beta$ the phase $\psi_{\alpha\beta}$ is identically zero, whereas when $\alpha \neq \beta$ it becomes irrelevant because of the presence of $e^{i(\xi_\beta - \xi_\alpha)}$, which takes random values on the unit circle. For this reason, whenever one is interested in the interference of two fields, only the modulus $r_{\alpha\beta}$ matters. Similarly, the phases $\{\psi_{\alpha\beta}\}$ are irrelevant when second-order correlation functions are considered, as in Ref.~\cite{Rigovacca_2016}.
However, when computing intensity correlations of order higher than two, in the presence of three or more sources, cyclic products can appear. These terms have the form $\bigbraket{\phi_1}{\phi_2} \bigbraket{\phi_2}{\phi_3} \bigbraket{\phi_3}{\phi_1}$, and their total phase $\psi = \psi_{12} +  \psi_{23} + \psi_{31}$ is not masked by the random fluctuations of $\{\xi_\alpha\}_{\alpha=1}^3$. Therefore, the phase $\psi$ carries physical meaning and can affect the interference of the fields. This additional phase $\psi$ plays exactly the same role as the ``triad'' phase identified and studied in Ref.~\cite{Alex_2017}, but its origin is completely classical. We will discuss in more detail the small differences between the two in the following, after having introduced the quantum framework which corresponds to the classical scenario considered here. 

In the remainder of this section we provide an explicit expression for the classical third-order intensity correlation function $G_3^{(\text{cl})}$. Its derivation is long but straightforward, as one only needs to keep track of all the terms that do not become zero when the averages over the random phases of the fields are performed. For example, by writing as $\av{I_\alpha}$ the average intensity emitted by the $\alpha$th source, the expectation value of a single output intensity can be expanded as 
\begin{equation}
\av{I_i} = \sum_{\alpha=1}^3 |U_{i\alpha}|^2 \av{I_\alpha},
\end{equation}
because the expectation value of \Eq{eq: classical fields overlap} is nonzero only when $\alpha = \beta$. With similar, although more involved, considerations it is possible to obtain the following expression:
\begin{align}\label{eq: generic classical}
G^{(\text{cl})}_3  &= 1 + \frac{\mathcal F_1 + \mathcal F_2 + \mathcal F_3}{\prod_{i=1}^3(\sum_{\beta=1}^{3} |U_{i\beta}|^2 \av{I_\beta})} \notag \\
&+\sum_{\alpha < \beta}^{3} r_{\alpha\beta}^2   
\sum_{i\neq j}^{3}\frac{ U_{i\alpha} U_{i\beta}^* U_{j\beta}U_{j\alpha}^* \av{I_\alpha}\av{ I_\beta}}{\left(\sum_{\gamma}^{3} |U_{i\gamma}|^2 \av{I_\gamma}\right)\left(\sum_{\delta}^{3} |U_{j\delta}|^2 \av{I_\delta}\right)} \notag\\
& +  \frac{2 \;\text{Re}\left[r_{12} r_{23} r_{31} e^{i\psi} \text{perm}(U \star U_{\Id,\pi}^*) \right]\prod_{\alpha=1}^{3}\av{I_\alpha}}{\prod_{i=1}^3(\sum_{\beta=1}^{3} |U_{i\beta}|^2 \av{I_\beta})},
\end{align}
where $\star$ represents the entrywise product, $\text{perm}(\cdot)$ evaluates the permanent, and $U_{\Id,\pi}$ is the matrix obtained from $U$ by permuting its columns according to the cyclic permutation $\pi = (1,2,3)$.
The terms labeled as $\mathcal F_1$, $\mathcal F_2$, and $\mathcal F_3$, instead, are nonzero only when the sources emit light with fluctuating intensities, i.e. according to distributions $\{p_\alpha(I_\alpha)\}_{\alpha=1}^3$ that are not delta functions. Explicitly, these terms can be written as
\begin{align}
\mathcal F_1 &= \sum_{\alpha=1}^{3} |U_{1\alpha}|^2|U_{2\alpha}|^2|U_{3\alpha}|^2 \left(\av{I_\alpha^3}-\av{I_\alpha}^3\right), \\
\mathcal F_2 &= \sum_{\alpha\neq \beta}^{3}\left(\av{I_\alpha^2}-\av{I_\alpha}^2\right)\av{I_\beta}
\Big[|U_{1\beta}|^2|U_{2\alpha}|^2|U_{3\alpha}|^2 + \text{c.p.}\Big], \\
\mathcal F_3 &=  \sum_{\alpha\neq \beta}^{3} 2 \; r_{\alpha\beta}^2 \left(\av{I_\alpha^2}-\av{I_\alpha}^2\right)\av{I_\beta}\cdot \notag\\  
&\qquad\qquad \text{Re}\Big[
|U_{1\alpha}|^2 U_{2\alpha}U_{2\beta}^*U_{3\beta}U_{3\alpha}^* + \text{c.p.}\Big], \label{def: classical extra rsquared}
\end{align}
where c.p. stands for the cyclic permutations of the indexes $1,2,3$. For example, given a function $f(i,j,k)$, one has
\begin{equation}
f(1,2,3) + \text{c.p.} = f(1,2,3) + f(2,3,1) + f(3,1,2). 
\end{equation}

In this paper we are mostly interested in the dependence of $G_3$ upon the complex overlaps $\{\bigbraket{\phi_\alpha}{\phi_\beta}\}_{\alpha\neq \beta}$. The right-hand side of \Eq{eq: generic classical} shows that $G_3^{(\text{cl})}$ consists of three components: one that is independent of those overlaps, one that depends on the squares $\{r_{\alpha\beta}^2\}_{\alpha\neq \beta}$, and one that depends on the product of all three overlaps [last fraction in \Eq{eq: generic classical}]. The classical triad phase appears in this last contribution. Furthermore, we note that the presence of fluctuations in the intensities of the sources not only shifts the value of $G_3^{(\text{cl})}$, but also affects its dependence upon the field distinguishabilities via \Eq{def: classical extra rsquared}, which depends on the squared moduli $\{r_{\alpha\beta}^2\}_{\alpha\neq\beta}$.

\subsection{Quantum scenario}
Let us consider the quantum counterpart of the scenario previously discussed: three independent quantum sources emit pulses of light that evolve through a linear optical interferometer and are subsequently detected. The field emitted by the $\alpha$th source, for $\alpha = 1,2,3$, can be characterized by the creation operator $\aop^\dagger_{\alpha;\tilde\phi_\alpha}$, where the first subscript corresponds to the spatial mode of the field, whereas the second one gives information on its frequency and polarization degrees of freedom. More precisely, $\aop^\dagger_{\alpha;\tilde\phi_\alpha}$ can be expanded as \cite{Loudon_1990,Tamma2015}
\begin{equation}\label{eq: creation in mode phi}
\aop^\dagger_{\alpha;\tilde\phi_\alpha}  = \sum_{\zeta = 1}^2 \int_0^\infty \text{d}\omega \, \tilde\phi_\alpha(\omega,\zeta) \aop^\dagger_{\alpha;\omega,\zeta},
\end{equation}
where $\aop^\dagger_{\alpha;\omega,\zeta}$ creates a monochromatic photon with frequency $\omega$ and polarization $\zeta$ in the $\alpha$th spatial mode, and obeys
\begin{equation}
[\aop_{\omega,\zeta},\aop^\dagger_{\omega^\prime,\zeta^\prime}] = \delta(\omega - \omega^\prime) \delta_{\zeta \zeta^\prime}.
\end{equation} 
In order for the canonical commutation relations $[\aop_{\alpha;\tilde\phi_\alpha}, \aop_{\alpha;\tilde\phi_\alpha}^\dagger] = 1$ to be satisfied for all $\alpha$, each function $\tilde\phi_\alpha(\omega,\zeta)$ needs to be normalized. It can, therefore, be represented by a unit vector in a Hilbert space isomorph to $L^2(\mathbb{R})\otimes \mathbb{C}^2$, with basis elements $\{\ket{\omega,\zeta}\}_{\omega,\zeta}$, i.e.
\begin{equation}
\bigket{\tilde\phi_\alpha} = \sum_{\zeta = 1}^2 \int \text{d}\omega \, \tilde\phi_\alpha(\omega,\zeta) \ket{\omega,\zeta}.
\end{equation}
Such mode vectors have previously been used to study the interference of partially distinguishable photons \cite{Tichy_PartialIndist,Shchesnovich2015,Bergou_2012,Sugimoto_2010}.

The quantum counterpart of classical light fields emitted with random phases is made of quantum states that are diagonal in the Fock basis, as it can be seen by averaging over a uniform distribution of phases. More precisely, we can write the state emitted by the $\alpha$th source as
\begin{equation}\label{eq: quantum state}
\hat\rho_\alpha = \sum_{n=0}^\infty \frac{q_\alpha(n)}{n!} \big(\aop^\dagger_{\alpha;\tilde{\phi}_\alpha}\big)^n\ketbra{0}{0}\big(\aop_{\alpha;\tilde{\phi}_\alpha}\big)^n,
\end{equation}
where $q_\alpha$ characterizes its photon-number statistics.
This is consistent with the classical scenario previously considered, as we now discuss. First, in both cases the vectors characterizing the frequency-polarization degrees of freedom of the light emitted by the sources are not allowed to vary from one pulse to another. Second, on average the emitted fields are invariant under a phase transformation which sends, for the classical and quantum cases respectively, $\phi_\alpha(\omega,\zeta) \to e^{i\theta_\alpha} \phi_\alpha(\omega,\zeta)$ and $\tilde \phi_\alpha (\omega,\zeta) \to e^{i\theta_\alpha} \tilde\phi_\alpha(\omega,\zeta)$. In the classical case this is because $\theta_\alpha$ can be absorbed within the random phase $\xi_\alpha$ in \Eq{def: classical field}. In the quantum case, the reason is that \Eq{eq: quantum state} is invariant under this phase transformation. Finally, notice that in the quantum case it is not necessary to explicitly take into account the possibility of dealing with stochastic sources that emit states with varying photon-number statistics, as this effect can be taken into account by suitably changing the distributions $\{q_\alpha\}_{\alpha=1}^3$.

As in the classical scenario, each quantum source is connected to one of the input ports of a three-mode linear optical interferometer, whose evolution can be fully described by a unitary matrix $U$, acting only on the spatial modes of the photons. In particular, for any given mode $\tilde\phi$, the annihilation operators representing the fields after the evolution can be written as combination of the input ones as
\begin{equation}\label{eq: b mode expansion}
\bop_{i;\tilde\phi} = \sum_{\alpha=1}^3 U_{i\alpha} \aop_{\alpha;\tilde\phi}.
\end{equation}
This relation allows us to write an expression for the operators $\{\hat{I}_i\}_{i=1}^3$ appearing in the definition of $G_3^{(\text{Q})}$ in \Eq{def: G3 I quantum}
\begin{equation}\label{eq: quantum intensity}
\hat{I}_i = \sum_{\zeta = 1}^2 \int_0^\infty \text{d}\omega \,\hbar\omega \; \bop^\dagger_{i;\omega,\zeta}\bop_{i;\omega,\zeta},
\end{equation}
which count the total energy received in each output spatial mode. The expectation values appearing in \Eq{def: G3 I quantum} then have to be interpreted as
\begin{equation}
\av{\hat I_i} = \Tr{\hat I_i \bigotimes_{\alpha=1}^3 \hat\rho_\alpha},
\end{equation}
and similarly for $\av{\prod_{i=1}^3 \hat I_i}$. In this last case, for simplicity it is convenient to consider the expectation value of the normally ordered product of the three output intensities. The result does not change because $\left[\bop_{i;\phi},\bop_{j;\phi^\prime}^\dagger\right] = 0$ for $i\neq j$, as it can be verified by exploiting \Eq{eq: b mode expansion} and the unitarity of $U$. For example, the use of the normally ordered expression is convenient in all those cases where the three-mode linear optical interferometer is affected by losses. Indeed, one way to take this into account is to add additional columns to the unitary matrix $U$, one for every input mode of the virtual beamsplitters modeling the losses. The advantage of the normal ordering is that only the entries $U_{i\alpha}$ connecting the three physical input and output modes explicitly appear in the expression of $G_3^{(\text{Q})}$.

The remainder of this section will be devoted to the derivation of an explicit expression for $G_3^{(\text{Q})}$. In order to do so, we will start by expanding $\hat{I_j}$ in a way that resembles the classical expression in \Eq{eq: classical fields overlap}. This will then allow us to recover the desired formula for $G_3^{(\text{Q})}$ by straightforwardly manipulating the classical result for $G_3^{(\text{cl})}$.
First of all, notice that in evaluating $G_3^{(\text{Q})}$ every annihilation operator $\bop_{i; \omega,\zeta}$ acts from the left on a state of the form
\begin{equation}
	\ket{\{n_\alpha\}_\alpha} = \prod_{\alpha=1}^{3} \left(\aop^\dagger_{\alpha;\tilde{\phi}_\alpha}\right)^{n_\alpha} \ket{0}.
\end{equation}
Together with the fact that
\begin{align}
\aop_{\alpha;\omega,\zeta} \left(\aop^\dagger_{\alpha;\tilde{\phi}_\alpha}\right)^{n_\alpha} \ket{0} 
&= \tilde\phi_\alpha(\omega,\zeta) n_\alpha \left(\aop^\dagger_{\alpha;\tilde{\phi}_\alpha}\right)^{n_\alpha-1} \ket{0} \notag \\
&= \tilde\phi_\alpha(\omega,\zeta) \aop_{\alpha;\tilde\phi_\alpha} \left(\aop^\dagger_{\alpha;\tilde{\phi}_\alpha}\right)^{n_\alpha} \ket{0},
\end{align}
we have
\begin{equation}
\bop_{i;\omega,\zeta} \ket{\{n_\alpha\}_\alpha} = \sum_{\alpha=1}^{3} U_{i\alpha} \tilde\phi_\alpha(\omega,\zeta) \aop_{\alpha;\tilde\phi_\alpha} \ket{\{n_\alpha\}_\alpha}.
\end{equation}
In turn, by exploiting \Eq{eq: b mode expansion} and \Eq{eq: quantum intensity} this implies that we can effectively substitute any $\hat{I}_j$ appearing in $G_3^{(\text{Q})}$ with
\begin{align}\label{eq: quantum I expansion}
\sum_{\alpha,\beta=1}^{3} U_{j\alpha}^* U_{j\beta}
\braket{\phi_\alpha}{\phi_\beta} \sqrt{\mathcal E_\alpha\mathcal E_\beta} \;\aop^\dagger_{\alpha;\tilde\phi_\alpha}\aop_{\beta;\tilde\phi_\beta},
\end{align}
where we identified $\ket{\phi_\alpha}$ with the vector having coordinates
\begin{equation}\label{def: relation between phis}
\braket{\omega,\zeta}{\phi_\alpha} = \sqrt{\frac{\hbar \omega}{\mathcal E_\alpha}}  \bigbraket{\omega,\zeta}{\tilde{\phi}_\alpha}.
\end{equation}
Here the normalization $\mathcal E_\alpha$ is the energy associated with $\tilde{\phi}_\alpha$:
\begin{equation}\label{def: mathcal E}
\mathcal E_\alpha = \sum_{\zeta=1}^2 \int_{0}^\infty  \rm{d}\omega \, \hbar \omega \, |\tilde{\phi}_\alpha(\omega,\zeta)|^2.
\end{equation}
The vectors $\{\ket{\phi_\alpha}_\alpha\}_\alpha$ are the same in the classical and quantum scenarios, as in both cases their overlaps characterize the distinguishabilities of the fields whenever intensity measurements are performed. We point out that most authors characterize the fields distinguishabilities via $\{\bigket{\tilde{\phi}_\alpha}\}_\alpha$ rather than via $\{\ket{\phi_\alpha}\}_\alpha$, but this is only due to the fact that they consider photon-number measurements rather than intensity measurements. We decided to focus on the latter in order to make a fair comparison with the classical scenario, where the concept of photon number is not defined. However, the two approaches are approximately equivalent in the experimentally relevant case where the photon bandwidths are much smaller than their mean frequencies [see \Eq{def: relation between phis}].  
As a consequence, the phase $\psi$ of the product $\bigbraket{\phi_1}{\phi_2} \bigbraket{\phi_2}{\phi_3} \bigbraket{\phi_3}{\phi_1}$ is the direct analogous of the triad phase studied in Ref.~\cite{Alex_2017}. Therefore, the appearance of a phase that must be considered together with the moduli $\{r_{\alpha\beta}\}_{\alpha\neq\beta}$ in order to fully characterize the distinguishability of three photons is an instance of more general phases that appear in the interference of three or more optical fields, whether or not they are quantized. 

It is possible to understand how the quantum and classical expressions for $G_3$ are related by comparing \Eq{eq: quantum I expansion} with \Eq{eq: classical fields overlap}. We remind the reader that the numerator of $G_3$ is obtained by taking the expectation value of the  (normal ordered) product of three such terms, corresponding to $j=1,2,3$.
First of all, we point out that in the classical and quantum cases the same combinations of indexes, running over the sources, lead to nonzero expectation values. This is because the phases $e^{i(\xi_\alpha-\xi_\beta)}$ in \Eq{eq: classical fields overlap} select the terms whose quantum counterparts in $\prod_{i=1}^3\hat{I}_i$ preserve the photon number of each source, which are the only relevant terms because of the diagonal structure of \Eq{eq: quantum state}.
Then, a one-to-one mapping can be easily seen in the simple situation where the quantum sources emit coherent states $\{\ket{\sqrt{A_\alpha} e^{i\xi_\alpha}}\}_\alpha$ with random phases $\{\xi_\alpha\}_\alpha$. As it can be expected, when this is the case the quantum expression reduces to the classical one with $I_\alpha = \mathcal E_\alpha A_\alpha$, because each annihilation operator in $\prod_{i=1}^3\hat{I}_i$ can be applied to an input coherent state. More generally, for input quantum states with arbitrary photon-number statistics any classical expectation value $\av{I_\alpha I_\beta I_\gamma}$ (with $\alpha,\beta,\gamma$ not necessarily different) is substituted in $G_3^{(\text{Q})}$ by $\av{:\nop_\alpha\nop_\beta\nop_\gamma:}$, where $:\hat{X}:$ represents the normally ordered form of the operator $\hat{X}$. We can thus move from the expression of $G_{3}^{(\text{cl})}$ to that of $G_{3}^{(\text{Q})}$ by performing the following substitutions: 
\begin{align}
\av{I_\alpha} &\leftrightarrow \mathcal E_\alpha \av{:\nop_\alpha:} = \mathcal E_\alpha \av{\nop_\alpha},\\
\av{I_\alpha^2} &\leftrightarrow \mathcal E_\alpha^2 \av{:\nop_\alpha^2:} = \mathcal E_\alpha^2 \left[\av{\nop_\alpha^2} - \av{\nop_\alpha}\right], \label{eq: 2nd commutation}\\
\av{I_\alpha^3} &\leftrightarrow \mathcal E_\alpha^3 \av{:\nop_\alpha^3:} = \mathcal E_\alpha^3 \left[\av{\nop_\alpha^3} - (3 \av{\nop_\alpha^2} - 2\av{\nop_\alpha})\right].  \label{eq: 3rd commutation}
\end{align}
Therefore, if we write as $\tilde G_3^{(\text{cl})}$ the expression formally obtained from $G_3^{(\text{cl})}$ by substituting each $I_\alpha$ with $\mathcal E_\alpha \nop_\alpha$, we are left with
\begin{align} \label{res: comparison G3}
&G_3^{(\text{Q})} = \tilde G^{(\text{cl})}_3  - \sum_{\alpha = 1}^{3} \frac{\mathcal E_\alpha^3 |U_{1\alpha}|^2 |U_{2\alpha}|^2 |U_{3\alpha}|^2 (3 \av{\nop_\alpha^2} - 2 \av{\nop_\alpha})}{\prod_{i=1}^3(\sum_{\beta=1}^{3} |U_{i\beta}|^2 \av{\nop_\beta} \mathcal E_\beta)} \notag\\
&- \sum_{\alpha\neq \beta}^3  \mathcal E_\alpha^2 \mathcal E_\beta\av{\nop_\alpha}\av{\nop_\beta}\frac{|U_{1\alpha}|^2  |U_{2\beta}|^2 |U_{3\beta}|^2   +  \text{c.p.}}{\prod_{i=1}^3\left(\sum_{\gamma}^{3} |U_{i\gamma}|^2 \av{\nop_\gamma}\mathcal E_\gamma\right)} \notag \\
& -\sum_{\alpha\neq \beta}^{3} 2 \; r_{\alpha\beta}^2 \mathcal E_\alpha^2 \mathcal E_\beta\av{\nop_\alpha}\av{\nop_\beta}
\frac{\text{Re}\Big[
|U_{1\alpha}|^2 U_{2\alpha}U_{2\beta}^*U_{3\beta}U_{3\alpha}^* + \text{c.p.}\Big]}{\prod_{i=1}^3\left(\sum_{\gamma}^{3} |U_{i\gamma}|^2 \av{\nop_\gamma}\mathcal E_\gamma\right)}.
\end{align}
Note that the noncommutativity of the annihilation and creation operators does not change the dependence of $G_3$ upon $\psi$, which remains the same in the classical and quantum cases.

\section{Revival in intensity correlations}\label{sec: revivals}
Two of the three new terms that appear in $G_3^{(\text{Q})}$ with respect to $G_3^{(\text{cl})}$ are independent of the distinguishability parameters of the fields, while the last one only depends on the pairwise squared overlaps $\{r_{\alpha\beta}\}_{\alpha,\beta}$.
This observation suggests a way of detecting nonclassicality signatures by means of the third-order intensity correlation function. If we choose a certain ``path'' in the space of the distinguishability parameters $\{r_{12}(\delta),r_{23}(\delta),r_{31}(\delta),\psi(\delta)\}$, and consider the dependence of $G_3$ upon $\delta$, the quantization of the fields not only shifts the function $G_3(\delta)$ by a constant amount, but can also change the ``shape'' of its graph. It is therefore possible to certify the nonclassicality of the sources whenever an experiment reveals a shape of $G_3(\delta)$ that is not compatible with the classical behavior of $G_3^{(\text{cl})}(\delta)$ under the same experimental conditions. 

\begin{figure}
	\centering
	\includegraphics[scale = 1.2]{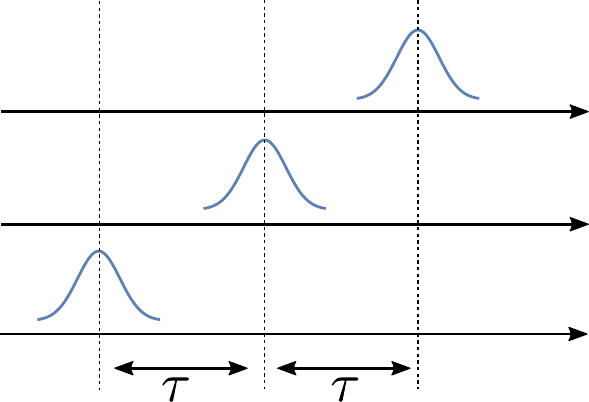}	
	\caption{(Color online) Time delays applied to three identical light pulses entering the three spatial modes of the linear optical interferometer. The parameter $\delta$ characterizing the distinguishability parameters of the fields is obtained by multiplying this delay $\tau$ with the spread in frequency $\sigma$ of the wave-packets.  \label{fig: shifted Gaussians}}
\end{figure}

In this paper we focus on a specific $\delta$-dependent path for the distinguishability parameters of the fields, which has been explored in recent experiments \cite{Spring_17,Tillmann2015,Rome_experiment,Alex_2017}. In particular, we consider three light fields with the same polarization and the same Gaussian wave-packet, i.e. such that the vector $\ket{\phi}$ describing their frequency and polarization has components
\begin{equation}
\braket{\omega,\zeta}{\phi} = \delta_{\zeta,\zeta_0} \frac{1}{(\sqrt{2\pi}\sigma)^{1/2}}e^{-\frac{(\omega-\omega_0)^2}{4\sigma^2}}, 
\end{equation}
where $\zeta_0$ is a certain fixed polarization.
Then, we apply to the fields emitted by the first and third sources respectively a time delay $-\tau$ and $+\tau$ with respect to the field emitted by the second one, as shown in Fig.~\ref{fig: shifted Gaussians}. In this way 
\begin{equation}\label{eq: modes}
\braket{\omega,\zeta}{\phi_1} e^{-i\omega \tau} = \braket{\omega,\zeta}{\phi_2} = \braket{\omega,\zeta}{\phi_3} e^{+i\omega \tau},
\end{equation}
so that the resulting distinguishability parameters are given by
\begin{equation}
\braket{\phi_1}{\phi_2} = \braket{\phi_2}{\phi_3} = e^{- i \omega_0 \tau} e^{-\frac{1}{2}(\omega\tau)^2},
\end{equation}
\begin{equation}
\braket{\phi_3}{\phi_1} = e^{2 i \omega_0 \tau} e^{-2(\omega\tau)^2},
\end{equation}
and the parameters relevant for $G_3$ become
\begin{equation}\label{def: parameters}
r_{12}(\delta) = r_{23} (\delta) = e^{-\frac{\delta^2}{2}}, \quad r_{31} (\delta) = e^{-2\delta^2}, \quad \psi(\delta) = 0,
\end{equation}
with $\delta = \sigma\tau$.

In the remainder of this section, we first provide an example in which the shape of $G_3(\delta)$ acts as an indicator of the nonclassicality of the sources. Then, in Sec.~\ref{sec: rev conditions} we develop some tools to classify the shape of $G_3(\delta)$, which can be used to certify the nonclassicality of the sources under less strict experimental conditions.

\subsection{Idea of revival in intensity correlations}\label{sec: ideal conditions}
Consider the evolution of three photons with the same energy, which evolve through a Bell three-port interferometer, characterized by the unitary matrix with entries
\begin{equation}\label{def: FT unitary}
U^{(\text{B})}_{j\alpha} \equiv e^{i \frac{2 \pi}{3}(j-1)(\alpha -1)}.
\end{equation}
In this case, the expression of $G^{(\text{Q})}_3$ given in \Eq{res: comparison G3} can be simplified as
\begin{equation}\label{res: quantum G3}
G_3^{(\text{Q})} = \frac{2}{9} - \frac{1}{9} (r_{12}^2 + r_{23}^2 + r_{31}^2)  + \frac{4}{9} r_{12}r_{23}r_{31} \cos{\psi}.
\end{equation}
Note that in this special case $G_3^{(\text{Q})}$ reduces to the probability of detecting a coincidence event, discussed in Ref.~\cite{Alex_2017}, because $\bigav{\hat{I}_1 \hat{I}_2 \hat{I}_3} \neq 0$ only when one photon is found per output mode. Therefore, the fact that even for indistinguishable photons $G_3 \neq 0$ implies that coincidence events are never completely suppressed, which is different from the Hong-Ou-Mandel result. This is a well known feature of Bell multiport interferometers with odd dimensions \cite{Lim_2005,Tichy2010}.

\begin{figure}
	\centering
	\includegraphics[scale = 0.6]{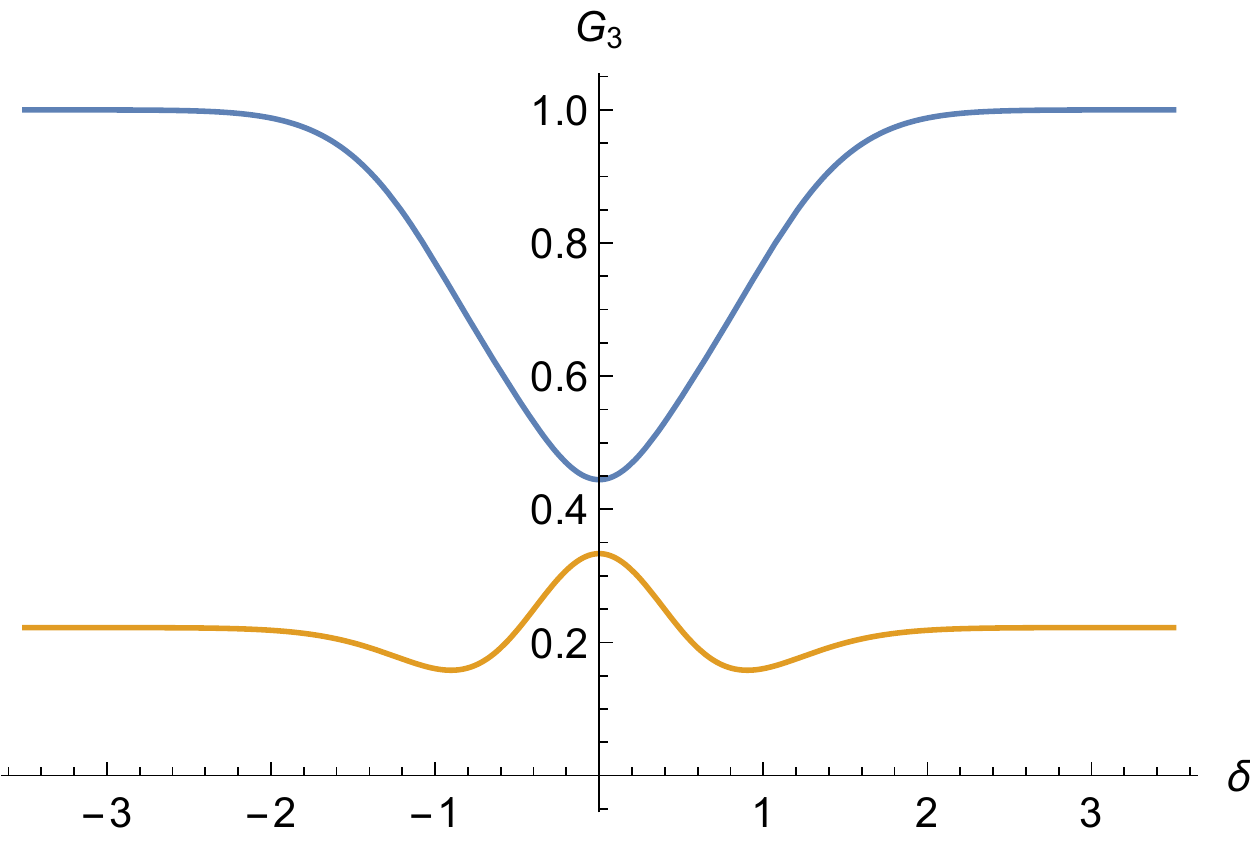}
	\caption{(Color online) Third order correlation function in the classical (top curve, blue) and quantum (bottom curve, orange) case, as a function of $\delta = \sigma \tau$. When $\delta = 0$, the light pulses reach the interferometer at the same time, and are indistinguishable.  \label{fig: sweep}}
\end{figure}

The classical counterpart of this scenario consists of three sources emitting pulses of light with the same intensities $I_\alpha \equiv I$. If the same linear optical setup is considered, the expression of $G_3^{(\text{cl})}$ given in \Eq{eq: generic classical} becomes
 \begin{equation}\label{res: classical G3}
 G_3^{(\text{cl})} = 1 - \frac{3}{9}( r_{12}^2 +  r_{23}^2 +  r_{31}^2) + \frac{4}{9}  r_{12}  r_{23}  r_{31} \cos{\psi}.
 \end{equation}
When the distinguishability parameters of the sources are chosen as in \Eq{def: parameters}, the corresponding functions $G_3^{(\text{cl})}(\delta)$ and $G_3^{(\text{Q})}(\delta)$ are characterized by the shapes shown in Fig.~\ref{fig: sweep}. Note that in the quantum case the positive contribution coming from the three-mode distinguishability term $ r_{12}r_{23}r_{31} \cos{\psi}$ is able to counteract the negative action of the pairwise distinguishabilities of the photons. As a consequence, a revival in the value of $G_3^{(\text{Q})}$ can be observed for small values of $\delta$, whereas this effect cannot be seen in the classical framework, where the negative coefficient multiplying $r_{12}^2 +  r_{23}^2 +  r_{31}^2$ is larger than in the quantum case.

This implies that if an experiment with sources satisfying \Eq{def: parameters} shows a revival in the third-order correlation function $G_3(\delta)$, the input fields cannot be explained by the chosen classical wave-like model of light, as long as the sources have the same intensities and the interferometer is characterized by \Eq{def: FT unitary}. In real experimental conditions, however, these additional requirements might not be exactly satisfied. For this reason it is important to understand when a revival in $G_3(\delta)$ cannot appear classically. Only in these cases the observation of a revival can be considered a nonclassicality signature. In order to perform these studies, however, we first need to find a way to characterize the presence or the absence of a revival in $G_3(\delta)$.

\subsection{Characterizing the presence or absence of revival} \label{sec: rev conditions}
In order to decide whether or not a certain function $G_3(\delta)$ shows a revival for $\delta \sim 0$, we can study the number of zeros of its first derivative. More precisely, a single zero corresponds to a regular dip, similar to the classical curve shown in Fig.\ref{fig: sweep}, whereas a revival similar to the quantum curve in that same figure appears if and only if $\partial_\delta G_3(\delta)$ has three zeros. 
We consider light sources whose distinguishability parameters can be written as in \Eq{def: parameters}, so that the third order correlation function $G_3$ can be expanded as
\begin{equation}\label{eq: structure G3}
G_3 = S - A e^{-\delta^2} - B e^{-4 \delta^2} + C e^{-3\delta^2},
\end{equation}
where the coefficients $S,A,B,C$ depend on the interferometer and on the sources.
Note that this structure appears not only in the classical case, but also in the quantum one, because the difference between the two scenarios only lies in the coefficients $S, A, B$. In particular, $C$ is the same in both cases, because the term proportional to $r_{12}r_{23}r_{31}\cos\psi = e^{-3 \delta^2}$ does not depend on the quantumness of the sources [see \Eq{res: comparison G3}]. 

It is convenient to write the first derivative of $G_3(\delta)$ as
\begin{equation}
\partial_\delta G_3(\delta) = \frac{2\delta}{e^{4\delta}} h\big(e^{\delta^2}\big),
\end{equation}
where $h$ is the polynomial function
\begin{equation}\label{der: h poly}
h(y) = A y^3 + 4 B - 3 C y.
\end{equation}
In this way, the number of zeros of $\partial_\delta G_3(\delta)$ depends only on the zeros of $h(y)$, which could be easily studied by considering $\partial_y h(y)$. More precisely, any value $y_0$ such that $h(y_0) = 0$ leads to a pair of zeros of $\partial_\delta G_3(\delta)$ (possibly degenerate) if and only if $y_0 \geq 1$, because $e^{\delta^2} \geq 1$ for all $\delta \in \mathbb{R}$. The absence of a revival in $G_3(\delta)$, therefore, corresponds to the absence of values $y_0>1$ such that $h(y_0) = 0$. On the other hand, a revival in $G_3(\delta)$ similar to the quantum curve in Fig.~\ref{fig: sweep} appears if and only if there exists a single $y_0 >1$ such that $h(y_0) = 0$.
We point out that in general the polynomial $h(y)$ might have more than one zero larger than 1, and when this happens the function $G_3(\delta)$ takes more complicated shapes.
On a case-by-case basis one should check which shapes of $G_3(\delta)$ could be obtained for classical sources under the experimental conditions of interest; the observation of any different result can then be considered a nonclassicality signature.

In this paper we focus on regimes such that the polynomial $h(y)$ is characterized by coefficients satisfying 
\begin{equation}\label{eq: conditions}
A,C > 0.
\end{equation}
This implies that $h(y)$ has no zeros $y_0 >1$ if and only if
\begin{equation}\label{cond: no rev}
h(y_{\min}) > 0 \qquad 
\vee 
\qquad
\begin{cases}
h(y_{\min}) \leq 0, \\
y_{\min} \leq 1, \\
h(1) \geq 0,
\end{cases}
\end{equation}
where $y_{\min}$ is the positive zero of $\partial_y h(y)$, i.e.
\begin{equation}\label{def: ymin}
y_{\min} = \sqrt{C/A}.
\end{equation}
Vice versa, when the conditions in \Eq{eq: conditions} hold, the polynomial $h(y)$ has at least one zero $y_0 >1$ if and only if
\begin{equation}\label{cond: rev}
\begin{cases}
h(y_{\min}) < 0, \\
y_{\min} > 1 \quad \vee \quad h(1)< 0.
\end{cases}
\end{equation}
These will be the main tools used in the following to decide whether or not a revival in $G_3(\delta)$ appears for a given interferometric setup and given classical or quantum sources.

\section{Classical No-Revival results} \label{sec: norev results}
Expanding on the preliminary considerations of Sec.~\ref{sec: revivals}, we now proceed to show that a revival in $G_3^{(\text{cl})}(\delta)$ cannot appear when classical fields interfere in a Bell three-port interferometer, irrespective of the probability distribution characterizing their intensities. After that, we numerically show that the same conclusion remains true if the interferometer only approximates the ideal Bell one. In doing this, we also take into account the effect of nonzero photon losses.

\subsection{Bell interferometer, generic classical sources} \label{sec: norev Bell int}
Consider a Bell three-port interferometer, and classical input light fields whose intensities are selected according to a probability distribution such that for all $\alpha = 1,2,3$
\begin{equation}\label{def: x and v}
\av{I_\alpha} \equiv x_\alpha, \quad \left[\av{I_\alpha^2} - \av{I_\alpha}^2\right] \equiv v_\alpha,
\end{equation}
for some nonnegative parameters $\{x_\alpha\}_{\alpha=1}^3$ and $\{v_\alpha\}_{\alpha=1}^3$, corresponding to the mean and variance. Notice that $\av{I_\alpha^3} - \av{I_\alpha}^3$ only appears within the $S$ coefficient in \Eq{eq: structure G3}, and for this reason it does not affect the shape of $G_3(\delta)$. Similarly, also all other higher-order central moments do not make a difference. Therefore, for the purposes of this section we can describe the classical sources only through the parameters in \Eq{def: x and v}.

Let us start by discussing the case in which $v_\alpha = 0$ for all $\alpha$, arising when the intensities emitted by the sources do not fluctuate from one pulse to another, though they may be different from each other. In this case, \Eq{eq: generic classical} can be written as in \Eq{eq: structure G3} with
\begin{align}
A &= 3 \,\frac{x_2(x_1 + x_3)}{(x_1 + x_2 + x_3)^2}, \label{eq: A QFTx}\\
B &= 3 \,\frac{x_1 x_3}{(x_1 + x_2 + x_3)^2}, \label{eq: B QFTx}\\
C &= 12 \,\frac{x_1 x_2 x_3}{(x_1 + x_2 + x_3)^3}, \label{eq: C QFTx}
\end{align}
which are all nonnegative. In order to show that no revival can be observed in $G_3^{(\text{cl})}(\delta)$ in this regime, it is sufficient to show that $y_{\min} \leq 1$ and $h(1)\geq 0$, because if this is the case \Eq{cond: no rev} is satisfied irrespective of the sign of $h(y_{\min})$. The inequality $y_{\min} \leq 1$ can be easily seen by defining for each $\alpha = 1,2,3$ the auxiliary variables   
\begin{equation}\label{eq: t subs}
t_\alpha = \frac{x_\alpha}{x_1+x_2+x_3},
\end{equation}
satisfying $t_1+t_2+t_3 =1$. This yields
\begin{equation}
y_{\min}^2 = \frac{4 t_1 t_3}{(t_1+t_3)}  \leq \frac{4 t_1 t_3}{(t_1+t_3)^2} \leq 1.
\end{equation}
Proving that $h(1) \geq 0$ is more involved, but it can be done in a similar way by introducing a Lagrange multiplier (see Appendix \ref{app: proof}).

If the variances $\{v_\alpha\}_{\alpha=1}^3$ of the input intensities are not identically zero, the coefficients $A^\prime, B^\prime, C^\prime$ characterizing $G_3^{(\text{cl})}(\delta)$ can be obtain from those in Eqs.~\eqref{eq: A QFTx}, \eqref{eq: B QFTx}, and \eqref{eq: C QFTx} as
\begin{align}
A^\prime &= A + 3 \, \frac{x_2(v_1+v_3) + v_2(x_1 + x_3)}{(x_1+x_2+x_3)^3}, \label{eq: Aprime}\\
B^\prime &= B + 3 \, \frac{x_1 v_3 + x_3 v_1}{(x_1+x_2+x_3)^3}, \label{eq: Bprime}\\
C^\prime &= C. \label{eq: Cprime}
\end{align}
Therefore, the difference between the polynomial $h^\prime(y)$ written in terms of $A^\prime$, $B^\prime$, $C^\prime$ and the previously studied polynomial $h(y)$ is nonnegative for all $y\geq 1$. As a consequence, if $h(y)$ had no zeros for $y\geq 1$, the same can be said for $h^\prime(y)$. No revival can thus appear in $G_3(\delta)$ for nonnegative $\{v_\alpha\}_{\alpha=1}^3$.

\subsection{Approximate Bell interferometers} \label{sec: norev numerics}

When different interferometric setups are considered, the complexity of the shapes appearing in $G_3^{(\text{cl})}(\delta)$ greatly increases. For example, for many randomly sampled interferometers a revival in $G_3^{(\text{cl})}(\delta)$ similar to that of Fig.~\ref{fig: sweep} can appear, when the classical sources have nonuniform intensities. However, the continuity of $G_3$ upon the evolution matrix characterizing the linear optical setup suggests that the behavior of $G_3^{(\text{cl})}(\delta)$ should not be too different from that in  Sec.~\ref{sec: norev Bell int} when the interferometer is ``close'' to the ideal case previously studied.
In this section we provide numerical evidences that this is indeed the case, thus greatly improving the applicability of our nonclassicality criterion.

\begin{figure}
	\centering
	\includegraphics{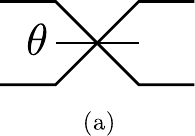}\hspace{2cm}
	\includegraphics{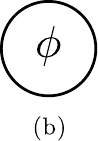} \\
	\vspace{0.5cm}
	\includegraphics{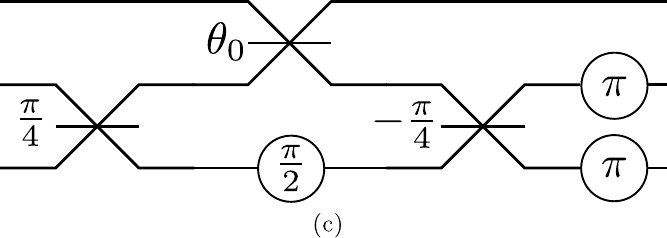}
	\caption{Panels $(a)$ and $(b)$ show, respectively, the symbols used to represent a beamsplitter $R_\theta$ and a phase shifter $P_\phi$, whose action on the input fields is described in the main text. The scheme used to decompose a Bell three-port interferometer into simpler components is shown in Panel $(c)$, where the light is assumed to enter from the left-hand side. Here $\theta_0 = \arccos(1/\sqrt{3})$. Notice that the two final phase shifts are required to reproduce the evolution matrix in \Eq{def: FT unitary}, but they are irrelevant for the purposes of measuring $G_3$, because only intensity measurements are performed. \label{fig: circuit}}
\end{figure}

In order to quantify the similarity between linear optical interferometers we take an operational approach. We start by decomposing the Bell three-port interferometer into a sequence of beamsplitters and phase shifters, as in Fig.~\ref{fig: circuit}. In particular, a beamsplitter acting on spatial modes $\alpha < \beta$ modifies the input fields through an evolution matrix 
\begin{equation}\label{def: BS}
R_\theta^{(\alpha,\beta)} = 
\left(\begin{array}{cc}
\cos\theta & \sin\theta \\
-\sin \theta & \cos\theta
\end{array}\right),
\end{equation}
which plays the same role of $U$ in \Eq{eq: classical interferometer}. The angle $\theta$ parametrizes the transmissivity of the optical element. Similarly, a phase shifter $P_\phi^{(\alpha)}$ acting on mode $\alpha$ applies a phase $e^{-i \phi}$ to the input field. With this notation, we have
\begin{equation}\label{eq: Bell decomposition}
U^{(B)} = P^{(2)}_\pi P^{(3)}_\pi R^{(2,3)}_{-\pi/4} R^{(1,2)}_{\theta_0} P^{(3)}_{\pi/2} R^{(2,3)}_{\pi/4},
\end{equation}
where $\theta_0 = \arccos\left(1/\sqrt{3}\right)$ (see Fig.~\ref{fig: circuit}).
Then, we add to the angle characterizing each element a certain degree of randomness, normally distributed with standard error $\epsilon$ around the ideal value that it would have in \Eq{eq: Bell decomposition}.
Finally, we include a nonzero amount of loss in every beamsplitter, quantified by a parameter $\eta \in [0,1]$ multiplying the evolution matrix on the right hand side of \Eq{def: BS}.

In our simulations we studied the behavior of $10^6$ graphs of $G_3^{(\text{cl})}(\delta)$, obtained by considering $\eta = 0.8$, $\epsilon = 2\pi/100$, and input intensities chosen at random. All of them had the shape of a regular dip, as the classical curve of Fig.~\ref{fig: sweep}. When we repeated the same analysis with $\epsilon = 2\pi \tfrac{3}{100}$, however, complex behaviors started to appear. Indeed, in a few occasions $G_3^{(\text{cl})}(\delta)$ became larger than its asymptotic value (given by $\delta \gg 1$), and in one case a revival could be observed for $\delta \sim 0$. From this analysis we can conclude that in order to interpret the observation of a revival in $G_3(\delta)$ as a nonclassicality signature, the angles characterizing the optical devices composing the interferometer should be at least within $2\pi/100$ from the ideal values reported in \Eq{eq: Bell decomposition}, with probability of losing each photon not higher than 36\% (i.e., $\eta \geq 0.8$). By considering parameters estimated in a specific setup of interest, the approach illustrated here can be adapted to provide a nonclassicality test in realistic settings.

\section{Origin of revival for quantum sources} \label{sec: nonclassicality}
When one moves from the classical to the quantum description of $G_3$, the main difference is given by the appearance of additional terms due to the commutation relations between annihilation and creation operators. The relevant equations showing this are \Eq{eq: 2nd commutation} and \Eq{eq: 3rd commutation}. As only $\av{I_\alpha^2}$ is multiplied by a function of $\delta$ in $G_3^{(\text{cl})}$, while $\av{I_\alpha^3}$ is not, for the purpose of understanding the change of shape between $G_3^{(\text{cl})}$ and $G_3^{(\text{Q})}$ we can limit our attention to \Eq{eq: 2nd commutation}. Effectively, the positive variance $\av{I_\alpha^2}- \av{I_\alpha}^2$ is substituted with the difference
\begin{equation}\label{eq: subPoiss difference}
\mathcal E_\alpha^2 \left[\av{\nop_\alpha^2}- \av{\nop_\alpha}^2 - \av{\nop_\alpha}\right],
\end{equation}
whose sign depends on the photon-number statistics of the input quantum state. Intuitively, we could have effects not obtainable by classical means only when \Eq{eq: subPoiss difference} becomes negative for some $\alpha$, i.e. when some source is characterized by a sub-Poissonian photon-number statistics. It will be convenient to express this property via the opposite of the Mandel Q parameter \cite{Mandel_79}
\begin{equation}\label{def: lambda parameter}
\mu_\alpha = - \frac{\av{\nop_\alpha^2}- \av{\nop_\alpha}^2 - \av{\nop_\alpha}}{\av{\nop_\alpha}^2},
\end{equation}
which is positive for sub-Poissonian sources and reaches its maximum value $1$ for single photons.

In the remainder of this section we show how a value of $\mu > 1/2$ is typically needed in order to observe a revival in $G_3^{(\text{Q})}(\delta)$. This is different from the nonclassicality criterion based on the second order intensity correlation functions developed in Ref.~\cite{Rigovacca_2016}, where any value of $\mu>0$ could lead to nonclassical results. For simplicity we consider a symmetric setup where the input quantum states are characterized by the same parameter $\mu_\alpha \equiv \mu \in [0,1]$.
At first we analytically study the quantum third-order correlation function obtained for a perfect Bell three-port interferometer, and then we numerically investigate the appearance of revivals when experimental imperfections are added to the picture.

\subsection{Bell three-port interferometer, quantum sources with the same photon-number statistics}
Exactly as in Sec.~\ref{sec: norev Bell int}, we can study the appearance of a revival in $G_3^{(\text{Q})}(\delta)$ by looking at the coefficients $A$, $B$, $C$ characterizing $\partial_\delta G_3^{(\text{Q})}(\delta)$ via the polynomial in \Eq{der: h poly}. 
At first we notice that the energies $\{\mathcal E_\alpha\}_\alpha$ can be simplified in the expression of $G_3^{(\text{Q})}$, because they are all the same [see Eqs.~\eqref{def: mathcal E} and \eqref{eq: modes}].
Therefore, by changing the definitions of $x_\alpha$ and $v_\alpha$ given in \Eq{def: x and v} into 
\begin{equation}
x_{\alpha} = \av{\nop_\alpha}, \qquad v_\alpha = \av{\nop_\alpha^2} - \av{\nop_\alpha}^2,
\end{equation}
we can straightforwardly obtain the coefficients $A$, $B$, $C$ characterizing $G_3^{(\text{Q})}$ by substituting each $v_\alpha$ with $v_\alpha - x_\alpha$ in 
Eqs.~\eqref{eq: Aprime}, \eqref{eq: Bprime} and \eqref{eq: Cprime}. For simplicity we focus on quantum states with the same sub-Poissonian photon-number statistics characterized by $\mu \in [0,1]$, so that we are left with
\begin{equation}
A = \frac{2}{9}(3-2\mu), \qquad
B = \frac{1}{9}(3-2\mu), \qquad
C = \frac{4}{9}.
\end{equation}
Note that $A,C >0$, so we can exploit the considerations made in Sec.~\ref{sec: rev conditions} in order to study for which $\mu$ a revival in $G_3^{(\text{Q})}(\delta)$ appears. A direct calculation shows that
\begin{equation}
y_{\min}^2 = \frac{2}{3-2\mu}, \qquad
h(1) = \frac{2}{3}(1-2\mu).
\end{equation}
This means that for $\mu \in [0,1/2]$ \Eq{cond: no rev} holds and there is no revival, while for $\mu \in ]1/2,1]$ a revival appears because \Eq{cond: rev} is satisfied (note that $y_{\min}>1$ implies $h(y_{\min}) < 0$ when $h(1)<0$).

Let us comment on this result. The presence of a revival for the maximum value of $\mu = 1$ corresponds to the single-photon case studied in Sec.~\ref{sec: ideal conditions}. We have now shown that a revival appears also for quantum sources with $\mu>\frac{1}{2}$. However, the nonclassicality contained in ``weak'' sub-Poissonian sources, i.e. with $\mu<1/2$, is not sufficient to achieve a revival. For example, the only Fock state with $\mu > 1/2$ is the single-photon state, because a two-photon state already has $\mu=1/2$.  

\subsection{Approximate Bell interferometers}
\begin{figure}
	\centering
	\includegraphics[scale = 0.653]{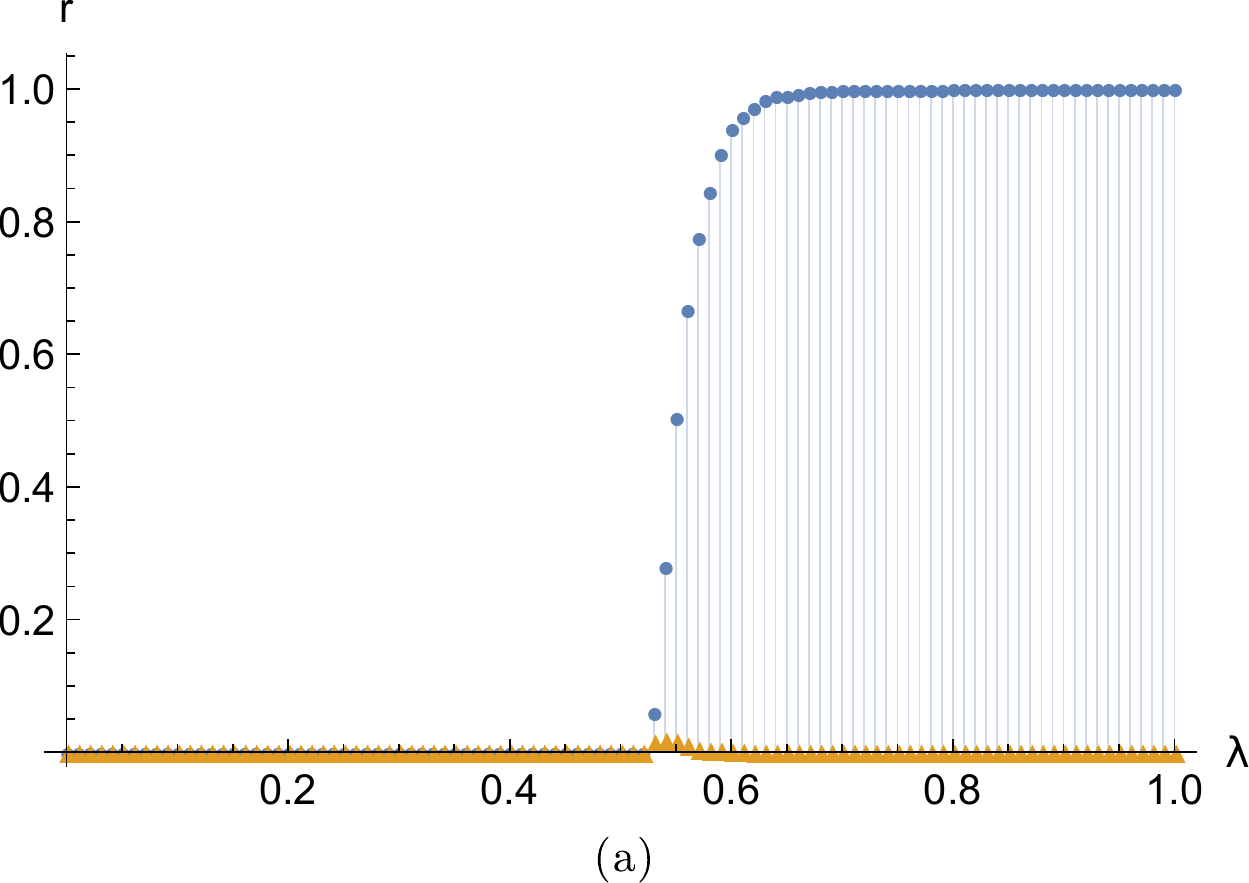}\\
	\vspace*{0.5cm}
	\includegraphics[scale = 0.9]{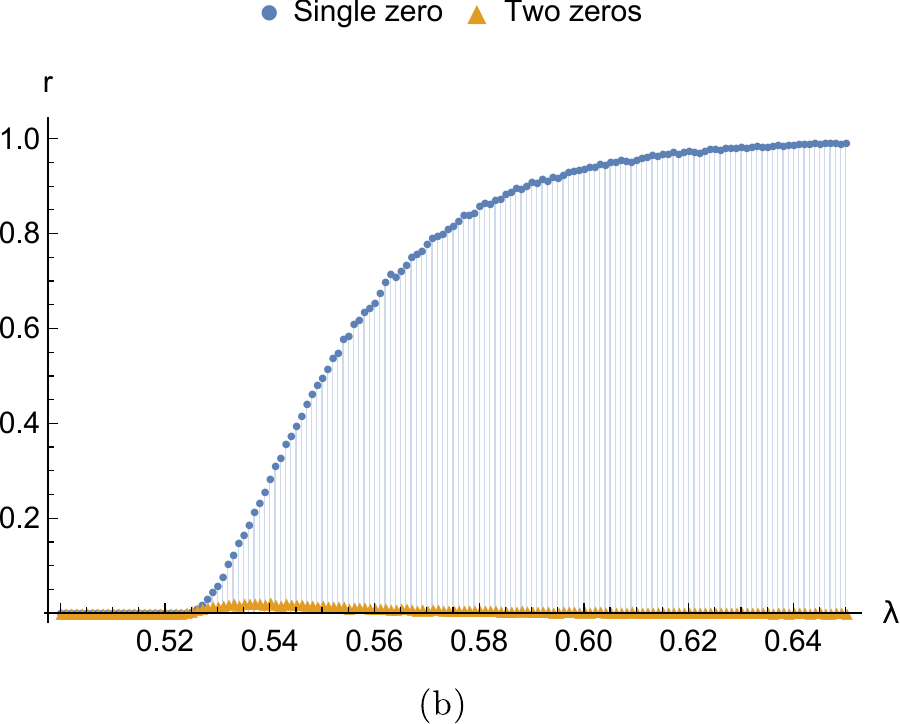}
	\caption{(Color online) Fraction $r$ of interferometric configurations leading to a revival of $G_3^{(\text{Q})}(\delta)$ against the sub-Poissonianity parameter $\mu$. Panel $(a)$ shows the full range {$\mu \in [0,1]$}, whereas Panel $(b)$ focuses on the region {$\mu \in [0.5,0.65]$}. Different markers are used to represent two possible revival ``shapes'', corresponding to the existence of one or two values $y_{1,2} > 1$ such that $h(y_{1,2}) = 0$. Details on the numerical simulation and comments on the existence of multiple revival shapes can be found in the main text.  \label{fig: quantum numerics}}
\end{figure}

We now consider imperfect interferometric setups, and we model them exactly as in Sec.~\ref{sec: norev numerics}, by slightly varying the parameters appearing in \Eq{eq: Bell decomposition}. We again consider three equivalent quantum states with photon-number statistics characterized by $\mu \in [0,1]$. By suitably applying time delays to the sources, we can recover the usual model for the distinguishability parameters of the fields, given in \Eq{def: parameters}. The shape of $G_3^{(\text{Q})}(\delta)$ can then be studied as a function of the sub-Poissonianity parameter $\mu$. More precisely, for any given value of $\mu$ we can randomly generate $10^4$ different imperfect Bell three-port interferometers obtained by setting the error $\epsilon = 2\pi/100$ and the loss parameter $\eta = 0.8$ (again as an example). Then we count how many times a revival appears, and in Fig.~\ref{fig: quantum numerics} we plot the fraction $r$ of revival appearances against the sub-Poissonianity parameter $\mu$. 

For $\mu = 0$ the photon-number statistics of the input states are Poissonian, implying that the sources emit phase averaged coherent states. This scenario is equivalent to a situation in which classical sources emit light fields with a random phase, and indeed we recover the same result of Sec.~\ref{sec: norev numerics}: no revival can be observed. The same behavior is maintained as long as $\mu\leq 1/2$, showing how weak sub-Poissonianity cannot lead to the appearance of revivals in $G_3(\delta)$ even for approximate Bell interferometers. The situation changes for $\mu>1/2$ (in particular for $\mu \gtrsim 0.52$, but this might be due to the relatively small size of the simulation, or to the choice of $\epsilon$). When $\mu$ goes above this threshold, revivals in $G_3(\delta)$ start to appear with a frequency that quickly rises to $100\%$. 

Interestingly, for a small range of values of $\mu$ it is also possible to observe more complex revivals (orange triangles in Fig.~\ref{fig: quantum numerics}), which corresponds to polynomials $h(y)$ having two zeros $y_{1,2}>1$. When this happens the shape of $G_3(\delta)$ becomes more complicated, and a small dip is added for $\delta\sim 0$ on top of the typical revival shown in Fig.~\ref{fig: sweep}. Despite this complexity, both types of revivals can be considered a quantum signature, because none of them appears in the corresponding classical setting (e.g., for $\mu = 0$, or for classical sources with possibly different intensities as those considered in Sec.~\ref{sec: norev numerics}).

\section{Discussion and conclusions} \label{sec: conclusion}
In this paper we considered the evolution of light through a three-mode linear optical interferometer, and we compared the prediction of quantum mechanics with those of a classical wave-like model of light. 
Two main assumptions have been made on the classical sources: independence and randomness of the phases characterizing the emitted light fields. The corresponding quantum states are described by a density matrix that is diagonal in the Fock basis. By considering only third-order correlations between output intensities, we were able to develop a criterion able to detect the nonclassicality of the input light, in the form of highly sub-Poissonian photon-number statistics. 

The idea behind our method is reminiscent of the Hong-Ou-Mandel effect between two identical, but time-delayed, input light fields. In that case, the second order correlation function $\av{I_1 I_2}/(\av{I_1}\av{I_2})$ is studied by sweeping the delay across a range of values around $0$. Depending on the depth of the observed dip, it is then possible to test the nonclassicality of the input fields (see e.g. Ref.~\cite{Rigovacca_2016}). In the method developed in this paper the sources need to be varied in a similar way, by applying opposite time delays to the first and third sources. Then, third-order correlations among output intensities are studied as a function of this delay. However, it is the shape of the obtained plot which provides information on the nonclassicality of the sources, rather than the minimum observed value of correlation.
Indeed, the dependence of $G_3$ on the distinguishability of the sources changes between the quantum and classical frameworks, because of the noncommutativity among annihilation and creation operators. The different interplays between the terms due to the distinguishability of two and three sources, then, can lead to the existence of shapes whose observation could not be possible in the presence of only classical fields.

In this paper we focus on a particular shape, motivated by recent experiments \cite{Spring_17,Tillmann2015,Rome_experiment,Alex_2017}: a revival of $G_3(\delta)$ for $\delta \sim 0$. We showed how for Bell three-port interferometers this shape cannot be observed classically, and that it can arise from identical quantum sources only when these are highly sub-Poissonian, as measured by the parameter $\mu$ in \Eq{def: lambda parameter}. As a consequence, our method can be used not only as an experimentally practical test of nonclassicality, but also as a certification of a high degree of sub-Poissonianity in quantum states. For example, this could be useful in assessing the quality of single-photon states.

At this stage there are questions still left unanswered. For example, it would be nice to formalize the notion of an interferometer ``close'' to a Bell one into a condition which, if satisfied, prevents a revival in $G_3^{(\text{cl})}$ from appearing.
As final remark, we point out that in a quantum framework $G_3$ is typically smaller than in a classical one (see e.g. Fig.~\ref{fig: sweep}). This is reminiscent of Ref.~\cite{Rigovacca_2016}, in which a classical tight lower bound was found for a second-order correlation function evaluated on the output intensities of a multiport interferometer. As this result holds for any possible linear optical interferometer and any set of classical sources with the properties described in Sec.~\ref{sec: classical scenario}, a violation of the aforementioned bound acts as a nonclassicality witness for the input light.
In a similar manner, it might be possible to find a tight lower bound on the values that $G_{3}^{(\text{cl})}$ can take, when evaluated for any possible three-mode interferometer. Then, the violation of such a bound would act as a nonclassicality witness exactly as in Ref.~\cite{Rigovacca_2016}. 
Due to the complexity of \Eq{eq: generic classical}, however, the minimization required to obtain the aforementioned lower bound was not completed in the present work. For example, we are aware of the fact that the symmetric experimental setup which in Ref.~\cite{Rigovacca_2016} led to the minimum amount of classical output intensity correlations is no more optimal for $G_3^{(\text{cl})}$. This investigation is left open for future research.

\section*{Acknowledgements} The authors warmly thank Alex E. Jones and Ian A. Walmsley for fruitful and inspiring discussions. This research was financially supported by the People Programme (Marie Curie Actions) of the EU's Seventh Framework Programme (FP7/2007-2013, REA Grant Agreement 317232) and by the EPSRC (Grant EP/K034480/1). C.D.F. acknowledges funding from the Singapore National Research Foundation (NRF-NRFF2016-02). M.S.K. thanks the Samsung GRO project and the Royal Society for support.

\bibliographystyle{apsrev4-1}
\bibliography{Qeffects}

%
%

\appendix

\section{Absence of classical revival for Bell three-port interferometers}\label{app: proof}
The sketch of the proof was already provided in the main text, here we just show that $h(1) \geq 0$, when the parameters $A$, $B$, $C$ are those in Eqs.~\eqref{eq: A QFTx}, \eqref{eq: B QFTx}, and \eqref{eq: C QFTx}. After the substitutions in \Eq{eq: t subs} the desired inequality can be written as
\begin{equation}\label{app: eq}
t_2(t_1 + t_3) + 4 t_1 t_3 - 12 t_1 t_2 t_3 \geq 0,
\end{equation}
subject to the constraint $t_1 + t_2 + t_3 = 1$.
We introduce a Lagrange multiplier $\lambda$, and set to zero the derivatives of 
\begin{equation}
\mathcal F = t_2(t_1 + t_3) + 4 t_1 t_3 - 12 t_1 t_2 t_3 - \lambda(t_1 + t_2 + t_3 - 1).
\end{equation}
This yields the following system of equations
\begin{align}
\frac{\partial \mathcal F}{\partial t_1} &= t_2 - 12 t_2 t_3 + 4t_3 - \lambda = 0, \label{app: first eq} \\
\frac{\partial \mathcal F}{\partial t_2} &= t_1 + t_3 - 12 t_1 t_3 - \lambda = 0, \\
\frac{\partial \mathcal F}{\partial t_3} &= t_2 - 12 t_2 t_1 + 4t_1 - \lambda = 0, \label{app: third eq}\\
\frac{\partial \mathcal F}{\partial \lambda} &=  1 - (t_1 + t_2 + t_3).
\end{align}
From \Eq{app: first eq} and \Eq{app: third eq} it follows that either $t_1 = t_3$ or $t_2 = 1/3$. The solutions $\bt = (t_1,t_2,t_3)$ are then respectively given by
\begin{align}
\bt^{(0)} &= \left(\frac{1}{6}, \frac{2}{3}, \frac{1}{6}\right), \\ \bt^{(1)} &= \left(\frac{1}{3}\left(1-\frac{\sqrt{3}}{2}\right), \frac{1}{3}, \frac{1}{3}\left(1+\frac{\sqrt{3}}{2}\right)\right), 
\end{align} 
which yield $h|_{\bt^{(0)}}(1) = 1/9$ and $h|_{\bt^{(1)}}(1) = 2/9$, both positive. In order to conclude that \Eq{app: eq} holds we should also check the positivity of the function on the left-hand side of \Eq{app: eq} on the border of the region defined by $t_1+t_2+t_3 = 1$ and $t_{\alpha} \geq 0$ for all $\alpha = 1,2,3$.
When $t_1 = 0$, $t_2 = 0$, or $t_3 = 0$, the left-hand side of \Eq{app: eq} reduces respectively to $t_2t_3$, $4t_1t_3$, or $t_1t_2$; these are all nonnegative functions, so the proof is concluded.

\end{document}